# Machine learning-based sampling of virtual experiments within the full stress state

A. Wessel[a,b], L. Morand[a], A. Butz[a], D. Helm[a], W. Volk[b]

[a] Fraunhofer Institute for Mechanics of Materials IWM, Woehlerstrasse 11, 79108 Freiburg, Germany

[b] Chair of Metal Forming and Casting, Technical University of Munich, Walther-Meissner-Strasse 4, 85748 Garching, Germany

E-mail address: alexander.wessel@iwm.fraunhofer.de

**Abstract**

This paper presents a new machine learning-based approach to investigate anisotropic yield surfaces of sheet metals by means of virtual experiments. The new sampling approach is based on the machine learning technique known as active learning, which has been adapted to efficiently sample virtual experiments with respect to the full stress state in order to identify parameters of anisotropic yield functions. The approach was employed to sample virtual experiments based on the crystal plasticity finite element method (CPFEM) for a DX56D deep drawing steel and compared with two state-of-the-art sampling methods taken from the literature. The resulting points on the initial yield surface for all three sampling methods were used to identify parameters of the anisotropic yield functions Hill48, Yld91, Yld2004-18p and Yld2004-27p. These parameters were then applied to a cylindrical cup drawing simulation to analyse the effect of the three sampling methods on a typical sheet forming simulation. The results show that the new machine learning-based sampling approach has a higher sampling efficiency than the two state-of-the-art sampling methods. Consequently, fewer computationally expensive crystal plasticity simulations are required. By comparing different variants of the Hill48, Yld91, Yld2004-18p and Yld2004-27p yield surfaces, it was also found that identifying parameters of anisotropic yield functions based on virtual experiments sampled within the full stress state can lead to a degraded representation of the in-plane anisotropy. With respect to DX56D deep drawing steel, this degradation was observed for the Yld2004-18p yield function. The negative implications following from this degraded in-plane representation were further demonstrated by the results of the cylindrical cup drawing process. As a consequence, the representation of the in-plane anisotropy must be carefully reviewed when taking the full stress state into account. In this context, Yld2004-27p was identified as being sufficiently flexible to simultaneously represent the plastic anisotropy of DX56D with respect to the in-plane and out-of-plane behaviour with high accuracy.

# 1. Introduction

Sheet metal forming operations play an important role in various manufacturing industries, particularly in the automotive sector. To reduce development times, minimise costs and increase the product quality of sheet metal parts, finite element simulations have become a state-of-the-art method to analyse and improve forming operations. One precondition for high-quality sheet metal forming simulations is an accurate description of the plastic material behaviour. Since sheet metal typically exhibits direction-dependent, or rather anisotropic material properties due to its manufacturing process, the mathematical description of texture-induced plastic anisotropy by anisotropic yield functions is essential (Banabic et al., 2010; Banabic et al., 2020; Tekkaya, 2000). Hence, many anisotropic yield functions have been developed for the plane and full stress state over the past decades. Apart from his well-known isotropic yield function (von Mises, 1913), von Mises (1928) also proposed the first anisotropic yield function for the plane and full stress state. It is a quadratic yield function and was initially introduced to describe the plastic anisotropy of single crystals. Using the concept of the plastic potential of von Mises (1928), Hill (1948) established a further anisotropic yield function, which can also be applied for the plane and full stress state. Hill's quadratic yield function has six material parameters for the full stress state and is typically used for body-centred cubic (bcc) materials such as steel (Vegter and Van den Boogaard, 2006). To increase the flexibility, Hill (1979, 1990, 1993) subsequently developed three anisotropic yield functions that focus on an enhanced representation of the anisotropic yield surface. Whereas the anisotropic yield functions proposed by Hill (1990, 1993) are designed for the plane stress state, the Hill (1979) yield function takes the full stress state into account.

A further important group of anisotropic yield functions that is relevant for the investigations in this study was developed by Barlat and co-workers. In the late 1980s, Barlat and Lian (1989) presented an anisotropic yield function for the plane stress state, which is often referred to as Yld89. This is based on the isotropic yield function introduced by Hershey (1954) and Hosford (1972) and has four material parameters. Two years later, Barlat et al. (1991) introduced the Yld91 yield function for the full stress state with six material parameters. The Yld91 yield function was then further developed, leading to the introduction of the two anisotropic yield functions Yld94 (Barlat et al., 1997) and Yld96 (Barlat et al., 1997). The number of material parameters amounts to six and eight respectively. In 2003, Barlat et al. (2003) also proposed the eight-parameter Yld2000-2d yield function for the plane stress state. Focussing on an enhanced representation of the plastic anisotropy for the full stress state, Barlat et al. (2005)

established the Yld2004-18p yield function. This is based on two linear transformations of the deviatoric stress tensor and contains 18 material parameters. Fourteen of these material parameters describe the in-plane behaviour, while the remaining four material parameters characterise the out-of-plane anisotropy. In addition, Van den Boogaard et al. (2016) demonstrated that the Yld2004-18p yield function has only twelve independent material parameters for the in-plane behaviour, and thus the number of independent parameters can be reduced to 16. Moreover, Aretz et al. (2010) introduced a modified version of the Yld2004-18p yield function by considering three linear transformations of the deviatoric stress tensor. Consequently, this anisotropic yield function has 27 material parameters and is referred to as Yld2004-27p. Of these 27 material parameters, 21 are related to the in-plane behaviour, while the remaining six parameters are associated with the out-of-plane behaviour. Additionally, Aretz and Barlat (2013) introduced two further anisotropic yield functions, called Yld2011-18p and Yld2011-27p. Both anisotropic yield functions have the same number of parameters as the Yld2004-18p and Yld2004-27p yield functions, but represent rather complementary, i.e. different-shaped, anisotropic yield surfaces. More recently, Lou et al. (2019) presented a modified version of the Yld2004-18p yield function with twelve material parameters that is suitable for materials with moderately plastic anisotropy. In this case, the number of parameters associated with the in-plane behaviour was reduced to eight, while the number of out-of-plane parameters remained constant.

Parameters of anisotropic yield functions are typically identified by various mechanical tests such as uniaxial tensile tests, hydraulic bulge tests or plane strain tension tests (Banabic et al., 2010; Banabic et al., 2020). To extend the data obtained experimentally, virtual experiments based on crystal plasticity (CP) simulations are widely used to identify parameters of anisotropic yield functions. With respect to single crystals, CP models to describe the deformation behaviour under external loading on the base of crystallographic slip were first developed by Peirce et al. (1982), Asaro (1983) and Peirce et al. (1983). However, the history of models for predicting the mechanical response of polycrystals goes back to the early 20$^{th}$ century, when Sachs (1929) proposed the first CP model assuming iso-stresses. Iso-stress means that the highest resolved stress acting on a slip system of a polycrystalline material is assumed for all grains within the polycrystal. In contrast to the iso-stress assumption, Taylor (1938) introduced a full-constraint (FC) CP model based on an iso-strain assumption. Later, this FC-Taylor model was enhanced by Bishop and Hill (1951) and renamed the Taylor-Bishop-Hill (TBH) model. Grain cluster models, like the LAMEL and the advanced LAMEL (ALAMEL) model by Van Houtte et al. (1999, 2005), take grain interaction within the polycrystal aggregate into account. Viscoplastic self-consistent (VPSC) models, which were first introduced by Moliniari et al. (1987) and later extended by Lebensohn and Tomé (1993, 1994), are homogenisation schemes that treat each grain of the polycrystal as an ellipsoidal inclusion embedded in a homogeneous equivalent medium. In addition to these mean field CP models, full field CP models directly incorporate microstructural information of the polycrystal such as the grain morphology. Full field CP models are typically employed in combination with the finite element method (FEM) or the fast Fourier transform (FFT). They were first used by Peirce et al. (1982) with a single crystal and then by several authors for a broad variety of material-related issues. An overview of different application examples appears in Roters et al. (2010). Nowadays, new types of CP models are emerging that use machine learning techniques such as neural networks to speed up CP simulations. Examples of machine learning-based CP models are presented in Yuan et al. (2018), Ali et al. (2019), Ibragimova et al. (2021), Bonatti et al. (2022), and Ibragimova et al. (2022).

There has been a steady increase in the amount of research into the use of virtual experiments to identify parameters of anisotropic yield functions. Barlat et al. (2005) first performed virtual experiments to identify parameters of anisotropic yield functions. Four virtual experiments using VPSC on AA2090-T3 and AA6111-T4 aluminium sheets were carried out to predict the resistance to shear relative to the thickness direction of the sheets, which cannot be measured experimentally. These results were then used to identify the out-of-plane parameters for the Yld2004-18p yield function. Similar approaches, where virtual experiments were performed as a substitute for real experiments, i.e. specific or rather experimentally realisable load cases were considered, were also conducted by Inal et al. (2010), Esmaeilpour et al. (2018), Han et al. (2020), Engler and Aretz (2021), Esmaeilpour et al. (2021), and Liu and Pang (2021).

Instead of replacing real experiments or carrying out virtual experiments with specifically defined load cases, further approaches can be found in the literature that focus on an enhanced exploration of the plane and full stress states. This means that a high number of virtual experiments is carried out by using a specific sampling method to explore the entire initial yield surface. For instance, Saai et al. (2013) varied the velocity at three master nodes of a representative volume element to sample virtual experiments with respect to the plane stress state. This approach was later also employed by Zhang et al. (2014, 2019). With respect to the full stress state, Grytten et al. (2008) sampled 690 virtual experiments based on a FC-Taylor model within the full stress state for AA5083-H116 aluminium sheets by considering a resolution of three points on the five axes of the five-dimensional strain rate space. Zhang et al.

(2015) performed 201 virtual experiments on AA1050 aluminium sheets using an extension of the Miller indices as suggested by Van Houtte et al. (2009) to identify parameters for the Yld2004-18p yield function. A random sampling approach was applied by Zhang et al. (2016) to perform 125 virtual experiments on a hot-band and a cold-rolled AA3104 aluminium sheet to calibrate parameters for the anisotropic yield functions Yld91, Yld2000-2d, Yld2004-18p and Yld2004-27p. Ma et al. (2022) also used a random sampling approach. They carried out a total of 60 virtual experiments for a DP980 dual phase steel to identify parameters for the Hill48, Yld91 and Yld2004-18p yield functions. A further example for using a random sampling approach in the context of virtual experiments was demonstrated by Nascimento et al. (2023). Here, 220 randomly sampled yield points were utilised to identify the parameters of a new neural network yield function.

Although different sampling methods are available in the literature, there has been little research on the efficiency of these sampling methods and their impact on the parameter identification of anisotropic yield functions for the full stress state. In a recent conference paper, Wessel et al. (2021) introduced a machine learning-based sampling approach for the plane stress state, which is based on the active learning technique "uncertainty sampling" (cf. Settles et al., 2012). A comparison with a random sampling approach demonstrated that the active learning-based sampling approach is advantageous in terms of sample efficiency and reliability with respect to the plane stress state when sampling virtual experiments. In addition, Qu et al. (2023) also applied an active learning strategy to improve the training of data-driven constitutive models for granular materials. The results demonstrated that data-driven constitutive modelling can benefit from active learning as fewer data are required. Furthermore, the importance of developing new computationally efficient methods to describe anisotropic yield surfaces based on virtual experiments, or rather CP simulations, is also underlined by the recent publications of Biswas et al. (2022), Fugh et al. (2022), and Schmidt et al. (2022).

Therefore, this study aims to improve the sampling of virtual experiments for the full stress state by introducing a new machine learning-based sampling approach. To this end, the concept of the machine learning-based sampling approach as introduced by Wessel et al. (2021) is extended to the full stress state and enhanced using the active learning strategy called "query by committee" introduced by Burbidge et al. (2007). Query by committee is a committee-based approach, which has already been applied to efficiently explore microstructure-property spaces in Morand et al. (2022), for example. In this work, the query by committee approach is applied to actively learn mapping from linear load paths to corresponding points on the initial yield surface. The performance of the new machine learning-based sampling approach is evaluated by comparing it with two state-of-the-art sampling methods taken from the literature. Moreover, this paper focusses on the effect of the sampling method on the parameter identification for anisotropic yield functions. To this end, all three sampling methods are used to identify parameters of the anisotropic yield functions Hill48, Yld91, Yld2004-18p and Yld2004-27p for a DX56D deep drawing steel. All anisotropic yield functions are also analysed regarding their capability to represent the plastic anisotropy with respect to the in-plane and out-of-plane behaviour. To make this comparison even more tangible, all anisotropic yield functions are applied to a cylindrical cup drawing simulation.

This article is organised as follows: While Section 1 serves as an introduction, Section 2 presents the materials and methods. This includes a detailed representation of the experimental characterisation methods for DX56D deep drawing steel as well as the crystal plasticity simulation framework used for performing virtual experiments within the full stress state. Particular attention is given to the advent of the new machine learning-based sampling method. The results are summarised in Section 3. Starting with the results of the experimental characterisation and the microstructure model, the results of the new machine learning-based sampling method are presented and compared with the two state-of-the-art sampling methods taken from the literature. Following the results of the three sampling methods, different variants of the anisotropic yield surfaces for Hill48, Yld91, Yld2004-18p and Yld2004-27p are compared with each other. This also includes an analysis of the in-plane anisotropy as well as the results of the cylindrical cup drawing simulations. The discussion is presented in Section 4 and focusses on the performance of the new machine learning-based sampling method as well as the effect of the different sampling methods on the parameter identification for anisotropic yield functions for the full stress state. The final conclusions are drawn in Section 5.

## 2. Materials and methods

This section outlines the materials and methods used in this study. First, the experimental methods for characterising DX56D deep drawing steel are summarised. Second, the crystal plasticity framework as well as the sampling methods for performing virtual experiments within the full stress state are explained. This includes details of the new machine learning-based sampling method as well as the two state-of-the-art sampling methods taken from the literature. Furthermore, information on the post-processing of the virtual experiments as well as the parameter identification for the anisotropic yield functions Hill48, Yld91, Yld2004-18p and Yld2004-27p are presented. As all anisotropic yield functions were applied to a cylindrical cup drawing simulation, further details of the corresponding finite element model are given at the end of this section.

2.1. Experimental procedures

Cold-rolled sheets of DX56D deep drawing steel were supplied by thyssenkrupp Steel Europe AG. The sheets were hot-dip zinc coated and had a thickness of 1.5 mm. Their nominal chemical composition is summarised in Table 1.

Table 1: Nominal chemical composition of DX56D deep drawing steel shown as maximum values in wt.% as declared by thyssenkrupp Steel Europe AG.

| C | Si | Mn | P | S | Ti |
|---|----|----|---|---|----|
| 0.12 | 0.50 | 0.60 | 0.10 | 0.045 | 0.30 |

The mechanical characterisation of DX56D deep drawing steel was carried out by means of two different experiments. First, uniaxial tensile tests at 0°, 15°, 30°, 45°, 60°, 75° and 90° with respect to the rolling direction (RD) were performed on a ZwickRoell Kappa 50 DS uniaxial testing machine. The specimens were manufactured by water jet cutting and had a gauge length of 80 mm and gauge width of 20 mm in accordance with DIN EN ISO 6892, test piece type 2. All uniaxial tensile tests were carried out until fracture using a constant engineering strain rate of 0.002 s$^{-1}$. During the experiment, the change in the gauge length was measured in the longitudinal and transverse directions of the specimen using two tactile extensometers. Three identical samples were tested for each direction. Second, hydraulic bulge tests with a die diameter of 110.8 mm were conducted on a ZwickRoell BUP 600 sheet metal testing machine. The tests were performed with a constant engineering strain rate of approximately 0.002 s$^{-1}$ following the procedure suggested by Jocham et al. (2017). In the experiment, the oil pressure was directly measured in the chamber by a pressure sensor, while the strain field was determined by a GOM Aramis digital image correlation (DIC) system. The measurement frequency of the DIC system was set to its maximum value of 40 Hz. Three hydraulic bulge tests were carried out in line with the uniaxial tensile tests. For post-processing, hydraulic bulge data were analysed in accordance with DIN EN ISO 16808 using GOM Aramis Professional 2017 software.

Crystallographic orientations and further microstructural information were obtained by electron backscatter diffraction (EBSD) measurements of the longitudinal and transverse cross-sections. Scans were conducted in a Zeiss Supra 40VP scanning electron microscope (SEM) equipped with an EDAX-TSL EBSD system and OIM Data Collect 5.31 software. An accelerating voltage of 20 kV was used to scan an area of 850 μm x 850 μm using a hexagonal grid with a step size of 1.5 μm. The EBSD data were analysed using the Matlab toolbox MTEX 5.1.1 (Bachmann et al., 2010). Only measurement points with a confidence index greater than 0.1 were considered for post-processing, as recommended by Field (1997). A misorientation of 5° was used for grain reconstruction and only grains with more than 10 measurement points were considered in the analysis.

2.2. Virtual experiments

2.2.1. Crystal plasticity model

Virtual experiments were performed using a CP constitutive model implemented in the commercial finite element software Abaqus/Standard. The phenomenological CP model used in this study is based on the work of Asaro (1983) and on the numerical framework presented in Kalidindi et al. (1992). The CP model was implemented in the finite element code through a UMAT user subroutine developed for the studies presented in Pagenkopf et al. (2016). The basis of the CP model is the multiplicative decomposition of the deformation gradient $\mathbf{F}$ into its elastic and plastic parts, assuming finite deformations according to the idea put forward by Kröner (Kröner, 1959; Lee and Liu, 1967):

$$\mathbf{F} = \mathbf{F}_e \mathbf{F}_p. \tag{1}$$

The elastic part of the deformation gradient $\mathbf{F}_e$ represents the reversible deformation behaviour of the crystal lattice, while the irreversible permanent response due to crystallographic slip is described by the plastic deformation gradient $\mathbf{F}_p$. The evolution of plastic deformation is defined by the plastic part of the velocity gradient $\mathbf{L}_\mathrm{p}$ and expressed as the sum of the shear rates $\dot{\gamma}^\alpha$ acting on every slip system $\alpha$:

$$\mathbf{L}_\mathrm{p} = \dot{\mathbf{F}}_\mathrm{p}\mathbf{F}_\mathrm{p}^{-1} = \sum_{\alpha=1}^{n} \dot{\gamma}^\alpha\, \mathbf{m}^\alpha \otimes \mathbf{n}^\alpha. \tag{2}$$

The unit vectors $\mathbf{m}^\alpha$ and $\mathbf{n}^\alpha$ are the slip direction and the slip plane normal of the slip system respectively. The parameter $n$ represents the total number of slip systems. In accordance with the literature (Asaro, 1983; Franciosi, 1983; Raphanel and Van Houtte, 1985; Raabe et al., 2005; Baiker et al., 2014), 24 slip systems, crystallographically called $\{110\}\langle 111\rangle$ and $\{112\}\langle 111\rangle$, are incorporated in the crystal plasticity model for body-centred cubic materials. The evolution of the plastic shear rate $\dot{\gamma}^\alpha$ in Eq. (2) is expressed by means of a phenomenological approach using the power law-type equation

$$\dot{\gamma}^\alpha = \dot{\gamma}_0 \left|\frac{\tau^\alpha}{\tau_\mathrm{c}^\alpha}\right|^{1/m} \mathrm{sign}(\tau^\alpha), \tag{3}$$

where $\dot{\gamma}_0$ is the reference shear rate, $\tau^\alpha$ is the resolved shear stress acting on a slip system $\alpha$ and $m$ is the rate sensitivity of slip. Unlike the approach suggested by Asaro (1983) and Kalidindi et al. (1992), the critical shear stress $\tau_\mathrm{c}^\alpha$ of a slip system $\alpha$ is described according to Lebensohn et al. (2007), Prakash and Lebensohn (2009), and Zhang et al. (2015) by

$$\dot{\tau}_\mathrm{c}^\alpha = \frac{\mathrm{d}\bar{\tau}^\alpha}{\mathrm{d}\Gamma} \sum_{\beta=1}^{n} q^{\alpha\beta} |\dot{\gamma}^\beta| \tag{4}$$

with the extended Voce type hardening law according to Tomé et al. (1984):

$$\bar{\tau}^\alpha = \tau_0 + (\tau_1 + \theta_1 \Gamma)\left[1 - \exp\left(-\frac{\Gamma \theta_0}{\tau_1}\right)\right]. \tag{5}$$

The quantities $\tau_0$, $\tau_1$, $\theta_0$ and $\theta_1$ are material-dependent parameters and are assumed to be identical for all slip systems. While $\tau_0$ and $\theta_0$ describe the initial yield stress and initial hardening rate in the grain respectively, the asymptotic hardening behaviour for large strains is characterised by $\tau_1$ and $\theta_1$. In Eq. (4) and (5), $\Gamma$ is the accumulated plastic shear strain over all slip systems $n$, which is expressed as

$$\Gamma = \int_0^t \sum_{\alpha=1}^{n} |\dot{\gamma}^\alpha| \mathrm{d}t. \tag{6}$$

Interaction between two different slip systems $\alpha$ and $\beta$ is incorporated by the interaction matrix $q^{\alpha\beta}$. The interaction matrix represents the latent hardening behaviour of a crystal and has the form:

$$q^{\alpha\beta} = \begin{bmatrix} 1 & q & \cdots & q \\ q & \ddots & \ddots & \vdots \\ \vdots & \ddots & \ddots & q \\ q & \cdots & q & 1 \end{bmatrix}. \tag{7}$$

The off-diagonal parameter $q$ defines the ratio of the latent to the self-hardening rate. In line with Zhang et al. (2015), the same latent to self-hardening rate ratio is considered for all slip systems.

2.2.2. Representative volume element

The representative volume element (RVE), which is assumed to be representative for DX56D deep drawing steel, was set up using the software package Neper 3.5.2 (Quey, 2011). A rectangular cuboid with a normalised edge length of 1.0 and a total of 1000 grains was considered for the RVE model setup. Grain elongation as obtained by EBSD measurements was taken into account as well. Additionally, the polycrystal was discretised by 40x40x40 eight-node hexahedral elements with linear shape functions and full integration (element C3D8 in Abaqus). To incorporate the crystallographic texture of DX56D deep drawing steel into the RVE, the experimental results of the EBSD measurements were used to derive the orientation density function (ODF). The ODF was reconstructed by considering 1000 orientations using the Matlab toolbox MTEX 5.1.1. Each of the crystallographic orientations obtained was randomly assigned to one grain of the RVE.

Crystal plasticity parameters, which were taken from the literature, are summarised in Table 2. In this context, the data for pure crystalline iron published by Haynes (2014) were used as elastic constants $C_{11}$, $C_{12}$ and $C_{44}$ for DX56D deep drawing steel. The parameter for the rate sensitivity of slip $m$, which commonly varies between 0.01 and 0.05 (Raabe et al., 2005; Zhang et al., 2014; Zhang et al., 2015; Zhang et al., 2016), was set to a relatively low value of 0.0125. This was done to minimise the strain rate dependency of the crystal plasticity model. The hardening parameters $\tau_0$, $\tau_1$, $\theta_0$ and $\theta_1$ were identified by a reverse engineering approach using the commercial software LS-OPT 6.0. Thus, the hardening parameters were determined by fitting the effective stress-strain curve in RD of the microstructural model to the corresponding experimental stress-strain curves. The experimental stress-strain curves, the normalised yield stresses and the r-values of the uniaxial tensile tests at 15°, 30°, 45°, 60°, 75° and 90°, as well as the results of the hydraulic bulge tests were then used to validate the hardening parameters. To this end, virtual experiments of the uniaxial tensile tests at 15°, 30°, 45°, 60°, 75° and 90° with respect to RD were carried out based on a texture rotation of the RVE. Since texture rotation does not account for the grain morphology incorporated in RVEs, it can lead to errors.

To assess this error, an additional uniaxial tensile test was performed at 90° with respect to RD by directly applying a load in the $y$-direction.

Table 2: Crystal plasticity parameters taken from the literature.

| CP parameter | Unit | Value | Reference |
| --- | --- | --- | --- |
| $C_{11}$ | GPa | 226 | Haynes (2014) |
| $C_{12}$ | GPa | 140 | Haynes (2014) |
| $C_{44}$ | GPa | 116 | Haynes (2014) |
| $\dot{\gamma}_0$ | - | 0.001 | Raabe et al. (2005), Zhang et al. (2015), Zhang et al. (2016) |
| $m$ | - | 0.0125 | Pagenkopf et al. (2016) |
| $q$ | - | 1.4 | Kocks (1970) |

The numerical homogenisation scheme introduced by Schmidt (2011) was used to derive macroscopic quantities of the RVE. Thus, periodic boundary conditions were applied to the RVE by introducing three auxiliary nodes – one for each pair of opposite faces. In addition, a fourth auxiliary node was implemented to impede rigid body rotations. The translational degrees of freedom of one arbitrary node of the RVE were fixed to prevent rigid body translations.

2.2.3. Machine learning-based sampling of virtual experiments

Active learning is a machine learning technique that can be understood as an interactive process to train machine learning models (Settles et al., 2012). This means that a machine learning model is trained on an initial data set and new data points are generated iteratively at locations at which the model's prediction quality is assumed to be worst. With virtual experiments, active learning is applied to sample proportional strain paths in order to obtain points on the anisotropic yield surface. To define these proportional strain paths with respect to the full strain space, the three-dimensional deformation history of the RVE is prescribed by using a six-dimensional strain vector ($\varepsilon_{xx}$, $\varepsilon_{yy}$, $\varepsilon_{zz}$, $\varepsilon_{yz}$, $\varepsilon_{xz}$, $\varepsilon_{xy}$). Assuming plastic incompressibility, this six-dimensional vector is transformed to five-dimensions as suggested by Van Houtte et al. (1992):

$$\begin{aligned}
\varepsilon_1 &= \frac{\sqrt{2}}{2}(\varepsilon_{xx} - \varepsilon_{yy}) \\
\varepsilon_2 &= \frac{\sqrt{6}}{2}(\varepsilon_{xx} + \varepsilon_{yy}) \\
\varepsilon_3 &= \sqrt{2}\varepsilon_{yz} \\
\varepsilon_4 &= \sqrt{2}\varepsilon_{xz} \\
\varepsilon_5 &= \sqrt{2}\varepsilon_{xy}
\end{aligned} \qquad (8)$$

The corresponding components of the original six-dimensional strain vector follow from the inversion of Eq. (8) and are:

$$\begin{aligned}
\varepsilon_{xx} &= \frac{\sqrt{6}\varepsilon_2 + 3\sqrt{2}\varepsilon_1}{6} \\
\varepsilon_{yy} &= \frac{\sqrt{6}\varepsilon_2 - 3\sqrt{2}\varepsilon_1}{6} \\
\varepsilon_{zz} &= -\sqrt{\frac{2}{3}}\varepsilon_2 \\
\varepsilon_{yz} &= \frac{2}{\sqrt{2}}\varepsilon_3 \\
\varepsilon_{xz} &= \frac{2}{\sqrt{2}}\varepsilon_4 \\
\varepsilon_{xy} &= \frac{2}{\sqrt{2}}\varepsilon_5
\end{aligned} \qquad (9)$$

This transformation is based on the introduction of a five-dimensional orthogonal base system and has already been used by Grytten et al. (2008) and Zhang et al. (2015) to perform virtual experiments with respect to the full stress state. The five-dimensional strain vector in Eq. (8) also serves as a starting point for the active learning-based sampling approach and the two state-of-the-art sampling methods presented hereinafter. With respect to the boundary conditions of the RVE, the five-dimensional strain vector is scaled according to a reference strain rate so as to ensure constant strain rates, and is then applied to the three auxiliary nodes of the RVE as a displacement gradient.

The active learning strategy query by committee is applied in this study. As the output space for sampling points on the initial yield surface is continuous, the approach introduced in Burbidge et al. (2007) for regression problems is deployed. Here, the input-output relation for the mapping to be learned is defined by

$$\sigma_i = f(\varepsilon_i), \qquad (10)$$

where $\varepsilon_i$ and $\sigma_i$ are the proportional strain path given by the five-dimensional strain vector in Eq. (8) and the corresponding point on the six-dimensional yield surface respectively. It is assumed that well-sampled data points for the mapping in Eq. (10) are also well suited to describe the anisotropic yield surface.

The general concept of the query by committee approach applied here is shown in Fig. 1. It is based on a committee of $n$ regression models that learn the input-output relation in Eq. (10). After the regression models are trained, the input space is searched for the location, i.e. a five-dimensional strain vector $\varepsilon_i^*$, at which the committee disagrees the most. The disagreement is measured by the variance $s^2$, which is expressed according to Krogh and Vedelsby (1995) as

$$s^2(\varepsilon_i) = \sum_{\eta=1}^{n} \left( \hat{\sigma}_i^{(\eta)}(\varepsilon_i) - \bar{\sigma}_i(\varepsilon_i) \right)^2. \tag{11}$$

Here, $\hat{\sigma}_i^{(\eta)}$ denotes the prediction of the $\eta^{\text{th}}$ regression model and $\bar{\sigma}_i$ describes the mean overall predictions. New strain vectors $\varepsilon_i^*$ are identified by solving the optimisation problem

$$\varepsilon_i^* = \underset{\varepsilon_i}{\operatorname{argmax}}\left(s^2(\varepsilon_i)\right). \tag{12}$$

To ensure that the norm of the strain vector equals unity, a soft constraint is incorporated into Eq. (12). Therefore, the optimisation problem yields

$$\varepsilon_i^* = \underset{\varepsilon_i}{\operatorname{argmax}}\left(s^2(\varepsilon_i) - W^*\left(1 - \operatorname{norm}(\varepsilon_i)\right)^2\right), \tag{13}$$

where $W^*$ is a factor to weight the soft constraint. After solving the optimisation problem in Eq. (13), a new virtual experiment is performed that considers the new strain vector $\varepsilon_i^*$ in order to determine the corresponding yield point $\sigma_i^*$. The new data tuple $(\varepsilon_i^*, \sigma_i^*)$ that is generated is then added to the training data set and the active learning loop is repeated.

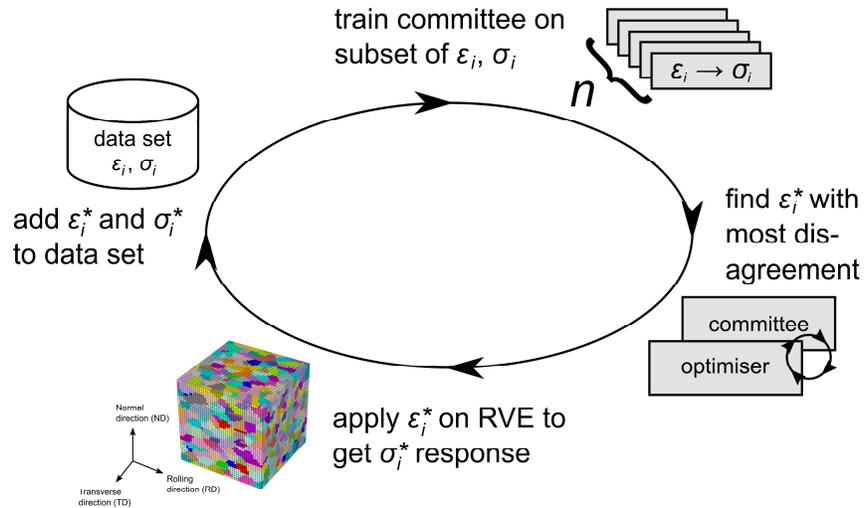

Fig. 1: Concept of the active learning-based sampling approach "query by committee" (cf. Morand et al., 2022) as adapted to perform virtual experiments.

In this study, the committee of regression models was realised by five neural networks, which were implemented using the Python package scikit-learn (Pedregosa et al., 2011). The neural networks consisted of three hidden layers with five, ten and ten neurons respectively. All use rectifiers as activation functions. The models were trained using the limited memory BFGS optimiser (Liu and Nocedal, 1989). Early stopping (Prechelt, 1998) was used for training, as well as an L2-regularisation (Krogh and Hertz, 1992) with a regularisation parameter of 0.00001. The differential evolution algorithm of Storn and Price (1997) as implemented in the Python package SciPy (Virtanen et al., 2020) was applied to solve the optimisation problem in Eq. (13) applying a weighting factor $W^*$ of 1000.

To assess the performance of the machine learning-based sampling approach, it was compared to two state-of-the-art sampling methods known from the literature. The first state-of-the-art method is a random sampling approach. Random sampling of virtual experiments was first done by Zhang et al. (2016) and subsequently employed by Ma et al. (2022) and Nascimento et al. (2023). The random sampling approach used in this study is based on the work of Muller (1959) and was originally introduced to generate uniformly distributed points on a hypersphere. Therefore, the five-dimensional strain vector described in Eq. (8) is defined by

$$\varepsilon_i = \frac{x_i}{\sqrt{x_1^2+x_2^2+x_3^2+x_4^2+x_5^2}}. \tag{14}$$

The values $x_i$ are drawn randomly from a Gaussian distribution with a mean of 0.0 and a standard deviation of 1.0 to sample five-dimensional strain vectors.

The second state-of-the-art sampling method was originally introduced by Van Houtte et al. (2009) and is based on an extension of the Miller indices. This means that the three Miller indices $[h_1, h_2, h_3]$ are extended to five dimensions $[h_1, h_2, h_3, h_4, h_5]$ and then used to define strain directions within the five-dimensional strain space. In this case, the five-dimensional strain vector in Eq. (8) is defined as:

$$\varepsilon_i = \frac{h_i}{\sqrt{h_1^2+h_2^2+h_3^2+h_4^2+h_5^2}}. \tag{15}$$

As suggested by Zhang et al. (2015), the following sets of five-dimensional Miller indices were considered: [0 0 0 0 1], [0 0 0 1 1], [0 0 1 1 1], [0 1 1 1 1], [1 1 1 1 1] and [1 1 1 1 3], including all permutations and changes of sign. This results in a total of 402 virtual experiments, which were carried out for the Miller indices-based sampling approach. Since the sequence of five-dimensional strain vectors defined using the active learning-based as well as the random sampling approaches is not the same with multiple repetitions, i.e. the sampling sequence is not unique, both sampling methods were repeated five times. Five individual sets of sampled points were thus investigated for both sampling methods. A total of 402 virtual experiments were performed for each of these sets so that they could be compared to the Miller indices-based sampling approach.

2.2.4. Parameter identification for anisotropic yield functions

Virtual experiments were evaluated using a specific plastic work per unit volume of 24.58 MPa, corresponding to a uniaxial true plastic strain of 0.1 in RD. This value was chosen to ensure the plastic anisotropy remained nearly constant, see Sections 3.1 and 3.2. The yield points were then used to identify parameters for anisotropic yield functions by minimising the error function:

$$E(\mathbf{c}) = \sum_{n=1}^{N} \left( \frac{\tilde{\sigma}(\sigma^n, \mathbf{c})}{\sigma_{ref}} - 1 \right)^2. \tag{16}$$

The variables $\boldsymbol{\sigma}^n$ and $\tilde{\sigma}$ are the stress tensor as predicted by a virtual experiment and the corresponding equivalent stress for a specific anisotropic yield function respectively. $N$ is the total number of yield points, and $\mathbf{c}$ denotes a vector containing the parameters of the anisotropic yield function under consideration. The flow stress of the virtual tensile test in RD was taken as the reference stress $\sigma_{ref}$. Eq. (16) was minimised by using the Sequential Least Squares Programming (SLSQP) algorithm as implemented in the Python package SciPy (Virtanen et al., 2020). The error function in Eq. (16) was also constrained so that the normalised equivalent stress in RD predicted by the anisotropic yield function is equal to 1. Using Eq. (16), in-plane and out-of-plane parameters of the anisotropic yield functions Hill48 (Hill, 1948), Yld91 (Barlat et al., 1991), Yld2004-18p (Barlat et al., 2005), and Yld2004-27p (Aretz et al., 2010) were identified. Since two parameters of the Yld2004-18p yield function are known to be dependent, the parameters $c'_{12}$ and $c'_{13}$ were set to unity as suggested by Van den Boogaard et al. (2016). Hence, only 16 parameters were taken into account for the parameter identification of Yld2004-18p.

To evaluate the anisotropic yield surfaces with respect to their in-plane behaviour, 100 additional virtual experiments applying a plane stress state were conducted by using the machine learning-based sampling method as originally introduced by Wessel et al. (2021). The resulting points on the yield surface were then utilised to identify the in-plane parameters of the Yld2004-18p and Yld2004-27p yield functions, as well as to calculate the root-mean-square deviations (RMSD) of the anisotropic yield surfaces representing the in-plane and out-of-plane anisotropy. Parameters associated with the out-of-plane behaviour were set to unity, which corresponds to the isotropic value. Again, as suggested by Van den Boogaard et al. (2016), the parameters $c'_{12}$ and $c'_{13}$ were set to unity so that only 12 independent parameters were considered for the in-plane version of Yld2004-18p.

2.3 Cup drawing simulation

To further analyse the parameters of the anisotropic yield functions identified, as well as to make the differences in the resulting anisotropic yield surfaces more tangible, all yield function parameters were applied to a cylindrical cup drawing simulation. The simulations were conducted by applying the commercial finite element software LS-DYNA with an explicit time integration scheme. The cup drawing process is schematically depicted in Fig. 2 and involves a deformable blank and three rigid tools. Due to the orthotropic material symmetry, only a quarter of the blank was considered in the analysis. The blank is meshed by 16 236 fully integrated continuum elements with a linear shape function (ELFORM = -2 in LS-DYNA). Three elements were applied over the blank thickness. Tools were modelled by Belytschko-Tsay shell elements (ELFORM = 2 in LS-DYNA).

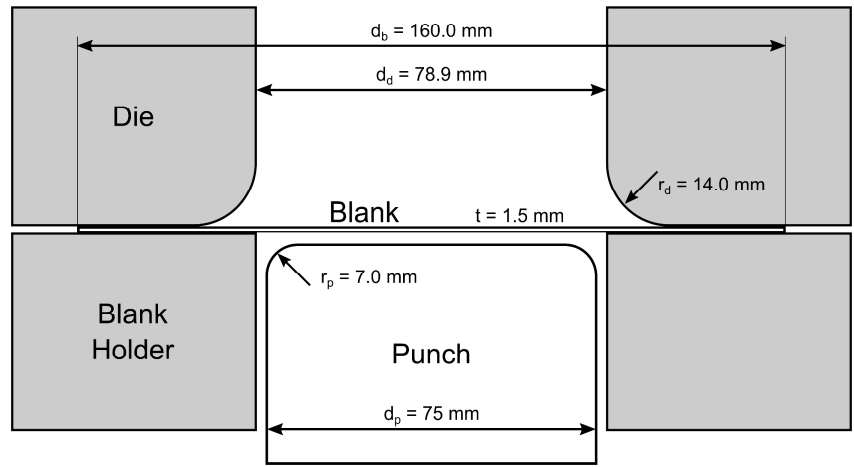

Fig. 2: Illustration of blank and tool geometries for the cylindrical cup drawing process.

The anisotropic yield functions were combined with an isotropic hardening law for the constitutive material model. To this end, the flow curve was determined by fitting the experimental results of the uniaxial tensile tests in RD and the hydraulic bulge tests to the modified Swift-Voce hardening law introduced by Kessler and Gerlach (2006). The well-known hardening laws of Voce (1948) and Swift (1954) are hereby merged via a linear combination. In the cup drawing simulations, the quarter of the blank is clamped by applying a blank holder force of 6.25 kN. All contact pairs were modelled with a constant Coulomb friction coefficient of 0.1, as performed by Yoon et al. (2006) for example.

## 3. Results

This section first presents the experimental results of the material characterisation for DX56D deep drawing steel. The crystallographic and mechanical results are then used to create the microstructure model as well as to validate its capability to represent the plastic anisotropy of DX56D deep drawing steel. The results of the virtual experiments obtained by means of the three sampling methods are then shown. This part includes an analysis of the sampling efficiency of the new active learning-based sampling approach as well as a comparison of all anisotropic yield surfaces determined by the three sampling methods. Subsequently, anisotropic yield surfaces for the full stress state are benchmarked with respect to the in-plane anisotropy. Finally, the earing profiles of the cylindrical cup drawing simulations are analysed.

3.1 Material characterisation

The representative stress-strain curves of the uniaxial tensile tests at 0°, 15°, 30°, 45°, 60°, 75° and 90° with respect to RD in Fig. 3 (a) show a direction-dependent plastic material behaviour for DX56D deep drawing steel. The stress-strain curves were highest at 45° and lowest at 90° with respect to RD. In addition, the results of the hydraulic bulge tests in Fig. 3 (b) indicate a high a level of formability with an average (± standard error) of 0.63 ± 0.02 for the maximum true strain before localisation.

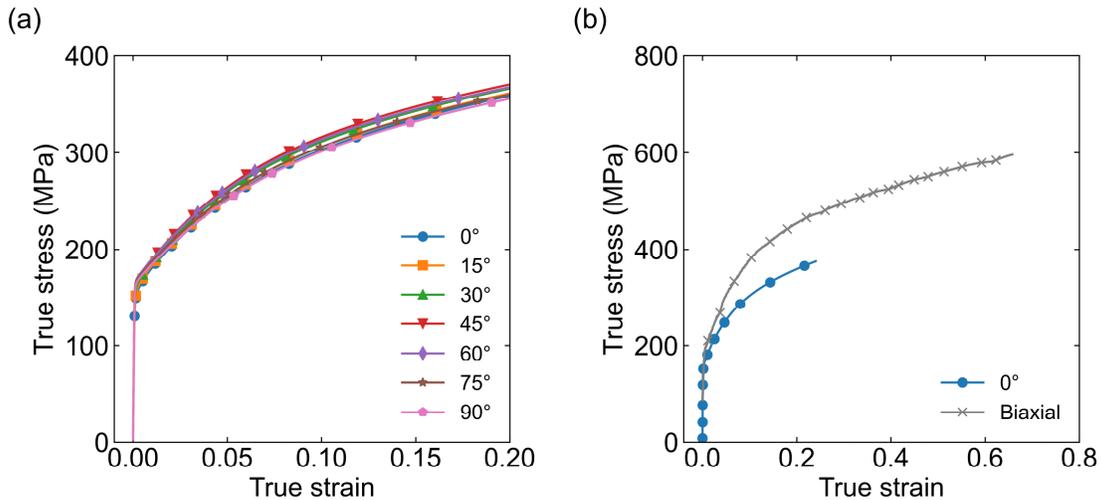

Fig. 3: Representative stress-strain curves of (a) uniaxial tensile tests at 0°, 15°, 30°, 45°, 60°, 75° and 90° with respect to the rolling direction (RD) and (b) hydraulic bulge tests representing a biaxial stress state for DX56D deep drawing steel. Only one of the three repetitions is illustrated by way of example.

The ratios of the normalised yield stresses in Fig. 4 illustrate that the plastic anisotropy of the hydraulic bulge test relative to the uniaxial tensile test at 0° with respect to RD is not constant during deformation, but changes continuously until a constant value is reached. In this case, normalised yield stresses determined by hydraulic bulge tests saturate at around 24.58 to 40.6 MPa specific plastic work, corresponding to an average true plastic strain in RD of 0.10 to 0.15 respectively. Conversely, normalised yield stresses of the uniaxial tensile tests at 15°, 30°, 45°, 60°, 75° and 90° with respect to RD remain constant from the beginning.

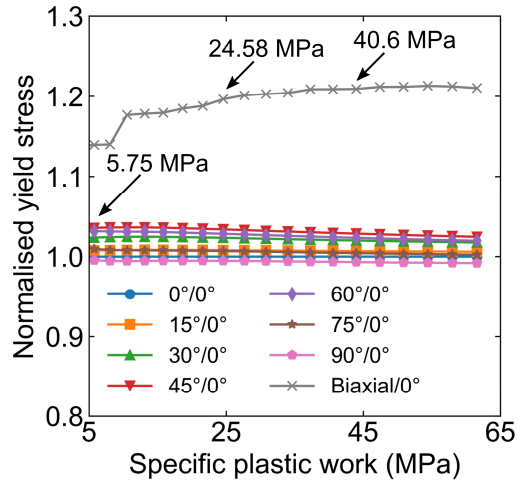

Fig. 4: Evaluation of the plastic anisotropy for DX56D deep drawing steel. Yield stresses were normalised by the average yield stress of the uniaxial tensile tests at 0° with respect to RD. Each data point represents the average of three repetitions. Due to a smooth elastic-plastic transition, the specific plastic work is illustrated from 5.75 MPa upwards.

Evaluated material properties of the uniaxial tensile tests and the hydraulic bulge tests are summarised in Table 3. Similar to the stress-strain curves in Fig. 3 (a), the results for the yield stress, yield strength and r-value demonstrate that DX56D deep drawing steel has a pronounced plastic anisotropy.

Table 3: Mechanical properties (mean ± standard error) of DX56D deep drawing steel obtained from uniaxial tensile tests in different directions with respect to RD, and hydraulic bulge tests representing a biaxial stress state.

| Direction | Flow stress[a] (MPa) | Yield strength[b] (MPa) | Uniform elong.[b] (%) | r-value[c] (-) |
|---|---|---|---|---|
| 0° | 303.7 ± 0.8 | 294.3 ± 0.9 | 27.0 ± 0.1 | 2.15 ± 0.04 |
| 15° | 306.1 ± 0.1 | 296.8 ± 0.2 | 26.2 ± 0.2 | 2.02 ± 0.02 |
| 30° | 310.8 ± 0.1 | 300.6 ± 0.1 | 25.3 ± 0.1 | 1.77 ± 0.01 |
| 45° | 314.0 ± 0.3 | 303.2 ± 0.4 | 24.4 ± 0.1 | 1.74 ± 0.02 |
| 60° | 312.3 ± 0.2 | 301.5 ± 0.2 | 24.8 ± 0.1 | 1.97 ± 0.02 |
| 75° | 305.8 ± 0.1 | 295.3 ± 0.2 | 25.6 ± 0.1 | 2.46 ± 0.01 |
| 90° | 302.1 ± 0.5 | 291.5 ± 0.6 | 26.2 ± 0.1 | 2.60 ± 0.05 |
| biaxial | 363.4 ± 0.2 | - | - | - |

[a] True stress at a specific plastic work of 24.58 MPa (corresponds to an average of 0.1 true plastic strain in RD)

[b] Engineering value

[c] Analysed between 0.1 and 0.2 true plastic strain according to DIN EN 10346

The inverse pole figure (IPF) maps of the longitudinal and transverse cross-sections in Fig. 5 show that <111> crystal directions tend to be preferably orientated parallel to the normal direction of the sheet metal and that DX56D thus has a pronounced $\gamma$-fibre. As to the grain morphology, both IPF maps contain a total of 5452 grains in the longitudinal and 5670 grains in the transverse cross-section. The grains are slightly elongated towards RD, with an aspect ratio of approximately 1.6. Their average size is roughly 133 $\mu m^2$ and 128 $\mu m^2$ in the longitudinal and the transverse direction respectively.

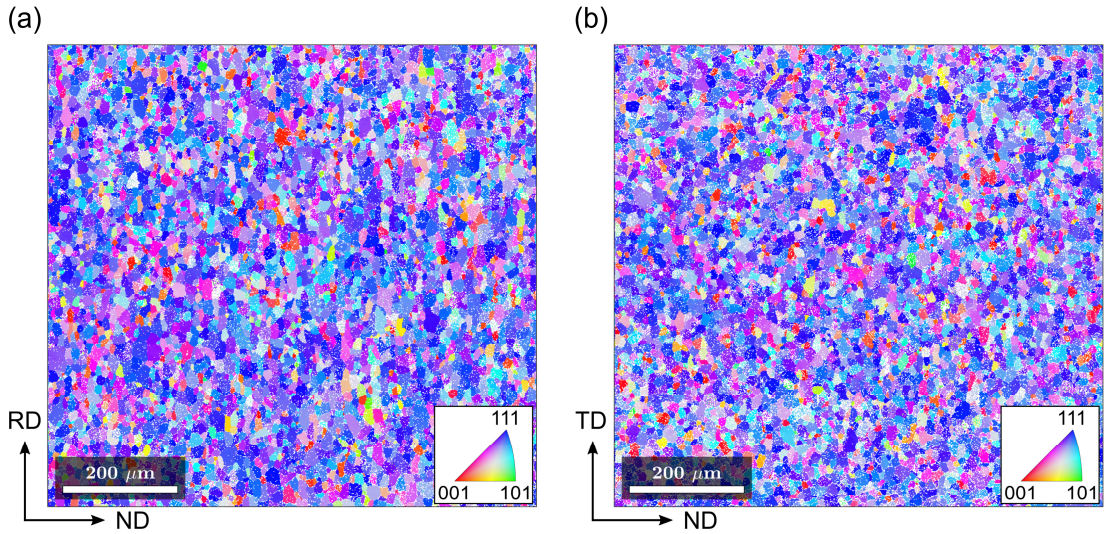

Fig. 5: Inverse pole figures (IPF) of the (a) longitudinal and (b) transverse cross-sections of DX56D deep drawing steel. IPFs were plotted with respect to the normal direction (ND).

Fig. 6 illustrates the ODF of DX56D deep drawing steel in the reduced Euler space with $0° \leq \varphi_1, \Phi, \varphi_2 \leq 90°$. As is the case in Fig. 5, a pronounced $\gamma$-fibre with a dominant crystallographic orientation $(111)\,[1\bar{2}1]$ is visible. The maximum intensity for the longitudinal and transverse cross-sections amounts to 11 multiples of a random density (MRD).

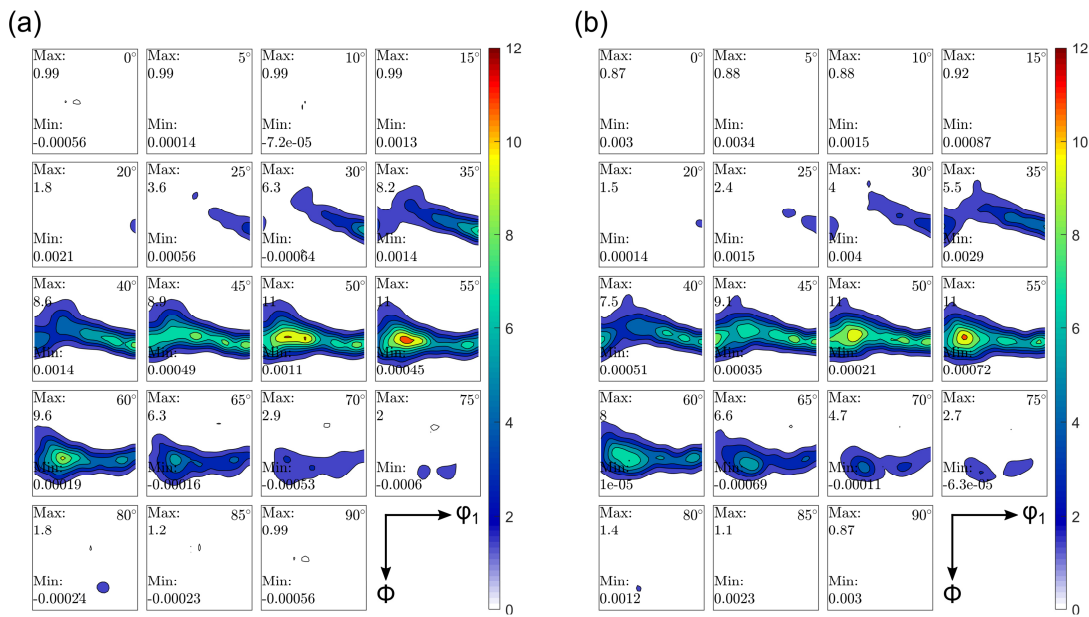

Fig. 6: Orientation density function (ODF) of DX56D deep drawing steel in the (a) longitudinal and (b) transverse cross-sections. Shown as $\varphi_2$-sections from 0° to 90° in steps of 5° through the reduced Euler space.

## 3.2 Microstructure model

The representative volume element generated for DX56D deep drawing steel is illustrated in Fig. 7 (a). Based on the results of the EBSD measurements in Section 3.1, grains were replicated by incorporating an elongation towards RD with an aspect ratio of 1.6.

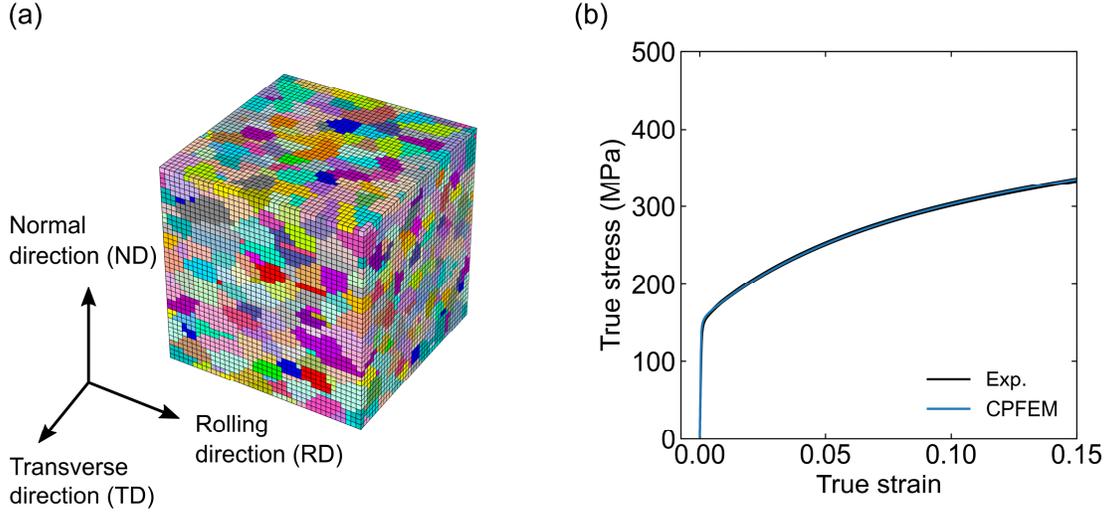

Fig. 7: (a) Microstructure of DX56D deep drawing steel shown as a representative volume element (RVE). Grains are elongated towards RD using an aspect ratio of 1.6 as determined by electron backscatter diffraction (EBSD) measurements. (b) Comparison of experimental (all three repetitions are illustrated) and stress-strain curves of a uniaxial tensile test in RD as predicted using a virtual experiment or rather the crystal plasticity finite element method (CPFEM).

The hardening parameters of the crystal plasticity model obtained using a reverse engineering approach are shown in Table 4. The corresponding stress-strain curve for the uniaxial tensile test in RD in Fig. 7 (b) correlates well with the experimental data.

Table 4: Hardening parameters of the crystal plasticity model identified for DX56D deep drawing steel using a reverse engineering approach. Values are given in MPa.

| $\tau_0$ | $\tau_1$ | $\theta_0$ | $\theta_1$ |
|---|---|---|---|
| 62.19 | 48.19 | 356.84 | 37.44 |

The results of the validation in Fig. 8 show that virtual experiments match the experimental normalised yield stresses and the r-values at 15°, 30°, 45°, 60°, 75° and 90° with respect to RD. However, compared to the reference solution obtained by direct loading in the $y$-direction, the virtual experiment based on a rotation of the texture leads to a lower normalised yield stress. The error amounts to roughly 1.6% with respect to the reference solution. In Fig. 8 (b), the texture rotation causes the r-value to be overestimated by approximately 4.7% with respect to the reference solution.

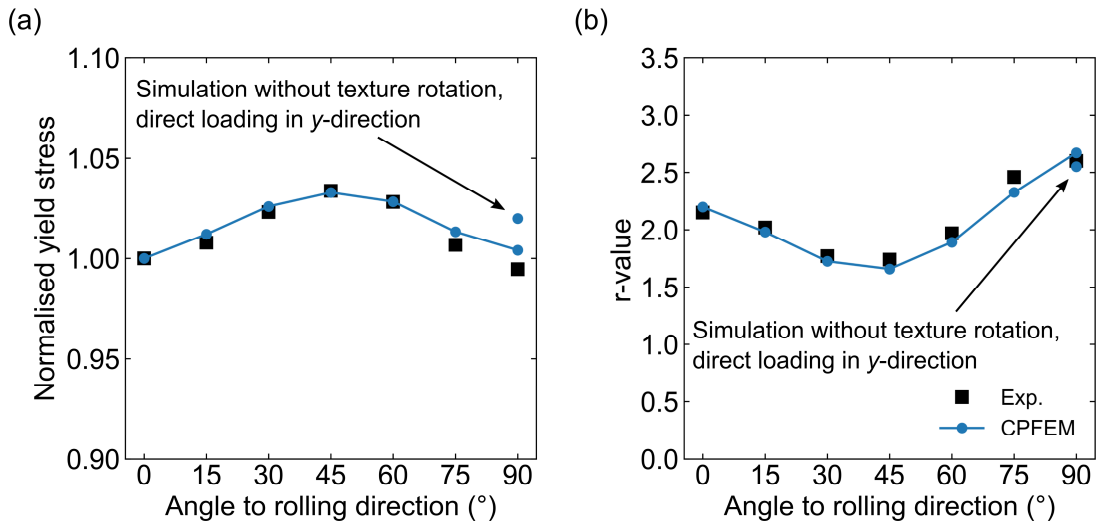

Fig. 8: Validation results of the (a) normalised yield stresses and (b) r-values with respect to RD. Yield stresses were determined using a specific plastic work of 24.58 MPa. Results in RD are not part of the validation as these results were adjusted by a reverse engineering approach to match the experimental data.

The validation results of the hydraulic bulge test in Fig. 9 (a) show a good match between the virtual experiment and the experimental data for true strains up to 0.1. The stress-strain curve as predicted by the virtual experiment slightly underestimates the experimental data for strains higher than 0.1. Furthermore, Fig. 9 (b) demonstrates that the microstructure model is able to reproduce the experimentally verified change in the biaxial yield stress in Fig. 4 at a slightly lower level. The maximum error between the two curves with respect to the experiment is approximately 3.4%.

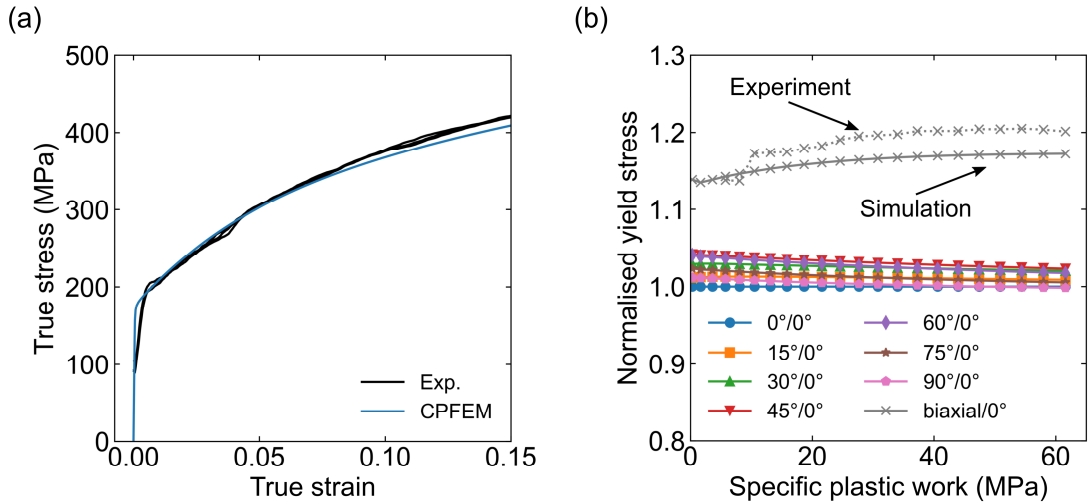

Fig. 9: a) Comparison of the experimental stress-strain curves obtained from hydraulic bulge tests (all three repetitions are illustrated) and as predicted by a virtual experiment. b) Normalised yield stresses obtained by crystal plasticity simulations at different levels of specific plastic work. Yield stresses were normalised by the uniaxial yield stress at 0° with respect to RD. The experimental results of the biaxial yield stress taken from Fig. 4 are also shown by way of comparison.

3.3 Sampling efficiency

As the Miller indices-based sampling approach is based on a fixed number of 402 virtual experiments, only the active learning-based and the random sampling approach were analysed with respect to their sampling efficiency, i.e. how many points on the yield surface are necessary. To this end, the fitting error between the virtual experiments of both sampling methods and the Hill48, Yld91, Yld2004-18p and Yld2004-27p yield surfaces was analysed in a three-step process: First, parameters of all anisotropic yield functions were identified for different numbers of yield points, i.e. 10, 20, …, 100, 125, …, 400, 402. Then, each of the parameters determined was used to calculate the respective error for the maximum number of 402 yield points, which served as a reference. Finally, the results were normalised by the error corresponding to a total of 402 yield points. Fig. 10 illustrates the evolution of this normalised error for the Hill48, Yld91, Yld2004-18p and Yld2004-27p yield functions. One can see that the normalised error of the Hill48, Yld91 and Yld2004-18p yield functions decreases faster for the active learning-based than the random sampling approaches. This means that parameters of these anisotropic yield functions – taking a deviation of 7.5% with respect to the reference state as a basis, for example – can be identified with even fewer virtual experiments. No significant differences between the active learning-based and the random sampling approaches are evident for Yld2004-27p. Additionally, the standard error, i.e. the variances between the five repetitions for the two sampling methods, is generally lower for the active learning-based approach than for the random sampling approach.

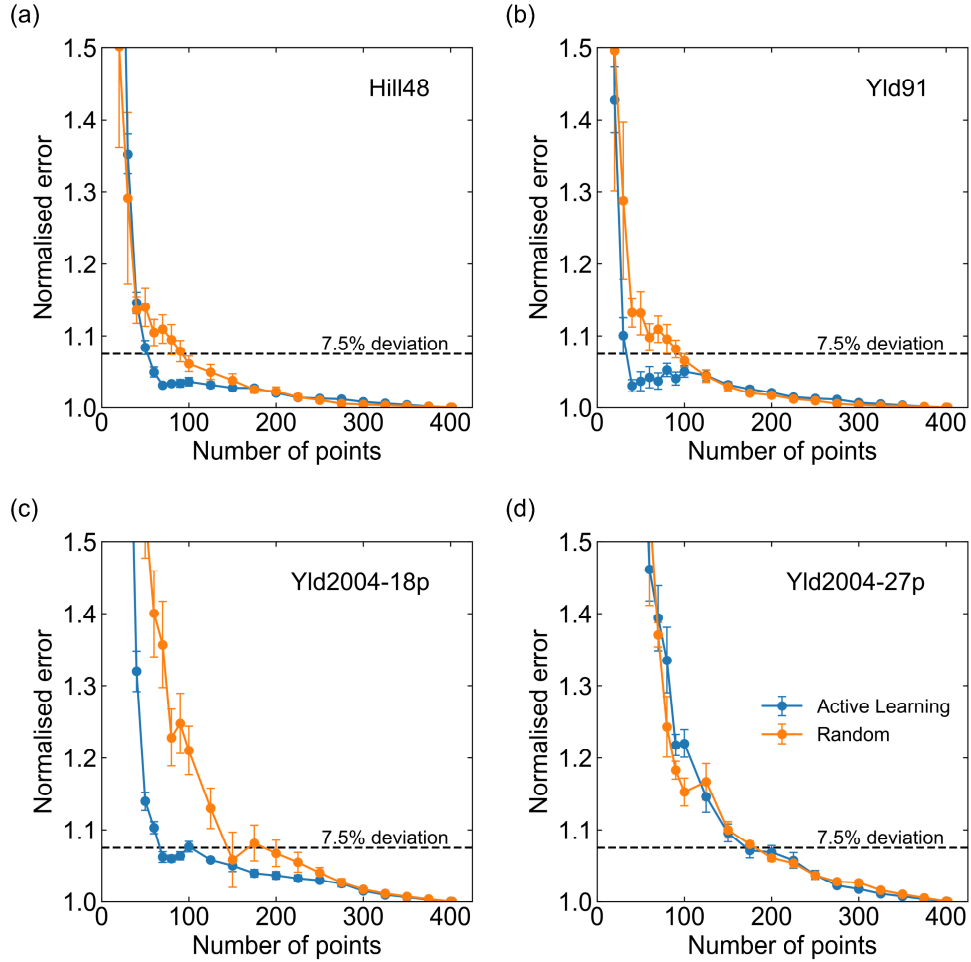

Fig. 10: Normalised error for the active learning-based and the random sampling approaches (mean ± standard error of 5 data sets in each case) in respect of the anisotropic yield functions (a) Hill48, (b) Yld91, (c) Yld2004-18p and (d) Yld2004-27p. To determine the normalised error, the error of each parameter set identified for a certain number of points ($n$ = 10, 20, … 100, 125, 400, 402) was calculated for the maximum of 402 yield points. Errors were subsequently normalised by the error corresponding to a total of 402 yield points.

For the subsequent analysis, parameters of the anisotropic yield functions Hill48, Yld91, Yld2004-18p and Yld2004-27p were identified at a specific number of yield points for the active learning-based and the random sampling approaches. To this end, the number of yield points corresponding to a deviation of less than 7.5% were chosen, as illustrated in Fig. 10. Table 5 summarises the resulting number of yield points used to identify the parameters of the anisotropic yield functions for the active learning-based and random sampling approaches.

Table 5: Number of yield points used to identify the parameters of Hill48, Yld91, Yld2004-18p and Yld2004-27p yield functions for the active learning-based and random sampling approaches.

| Sampling method | Data set | Hill48 | Yld91 | Yld2004-18p | Yld2004-27p |
|---|---|---|---|---|---|
| Active learning | 1 – 5 | 60 | 40 | 125 | 200 |
| Random | 1 – 5 | 100 | 100 | 200 | 200 |

Anisotropic yield surfaces as identified by the number of yield points in Table 5 and the maximum number of 402 yield points are compared in Fig. 11 for the Yld2004-27p yield function, using the active learning-based sampling approach. In general, the anisotropic yield surface of Yld2004-27p as identified at 200 yield points correlates well with the reference state of 402 yield points.

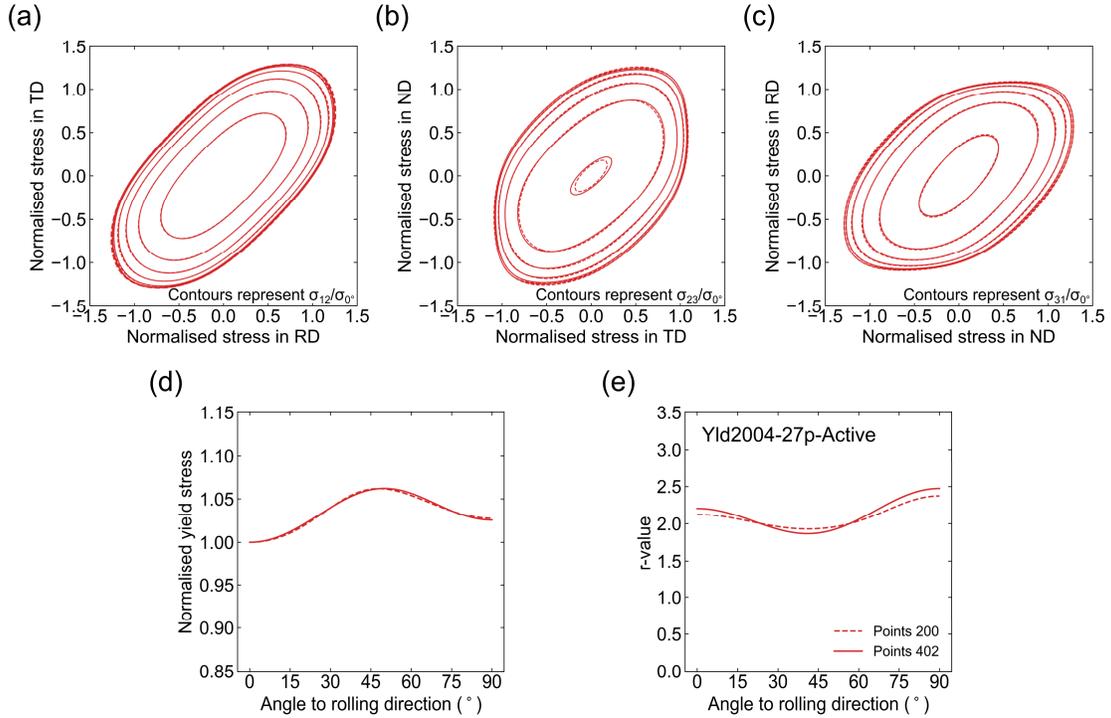

Fig. 11: Anisotropic yield surface for the Yld2004-27p yield function as identified by 200 and 402 yield points from data set 5 of the active learning-based sampling approach. Representation of the normalised yield surface with respect to (a) the RD-TD plane, (b) the TD-ND plane, (c) the ND-RD plane, and (d) the normalised yield stresses as well as (e) the r-values with respect to RD. Normalised shear contours shown in increments of 0.1 from 0.0 to 0.5.

### 3.4 Anisotropic yield surfaces

To assess the effect of the three sampling methods on the parameter identification of the four anisotropic yield functions for the full stress state, Fig. 12 and Fig. 13 compare the normalised yield surfaces, the normalised yield stresses and the r-values of all three sampling methods. Yield surfaces of the active learning-based and the random sampling approaches were identified according to Table 5 and averaged by computing the arithmetic mean of the five individual yield surfaces. Since the Miller indices-based sampling approach is based on a predefined number of 402 virtual experiments, all points had to be used to identify the parameters for the Hill48, Yld91, Yld2004-18p and Yld2004-27p yield functions. The comparison in Fig. 12 and Fig. 13 demonstrates that all three sampling methods lead to similar yield surfaces for the Hill48, Yld91, Yld2004-18p and Yld2004-27p yield functions. In fact, the differences between the four anisotropic yield functions, e.g. Hill48-Miller, Yld91-Miller, Yld2004-18p-Miller and Yld2004-27p-Miller, are greater than the differences between the three different sampling methods, e.g. Yld2004-27p-Active, Yld2004-27p-Miller and Yld2004-27p-Random.

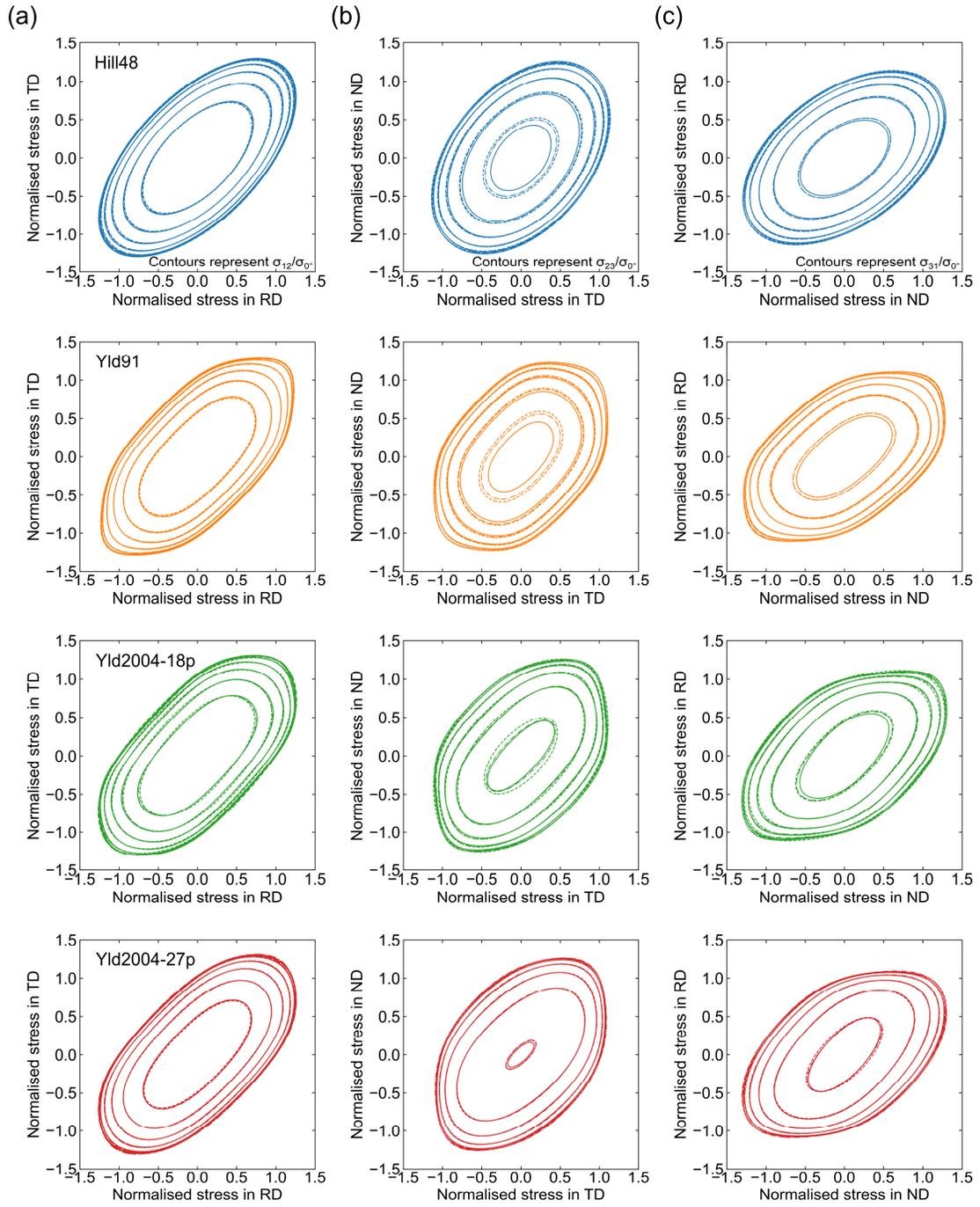

Fig. 12: Hill48, Yld91, Yld2004-18p and Yld2004-27p yield surfaces as identified by the active learning-based (solid line), random (dotted line) and Miller indices-based (dash-dotted line) sampling approaches with respect to (a) the RD-TD plane, (b) the TD-ND plane as well as (c) the ND-RD plane. Normalised shear contours shown in increments of 0.1 from 0.0 to 0.5. Yield surfaces identified by active learning-based and random sampling approaches were averaged by calculating the arithmetic mean of the five data sets.

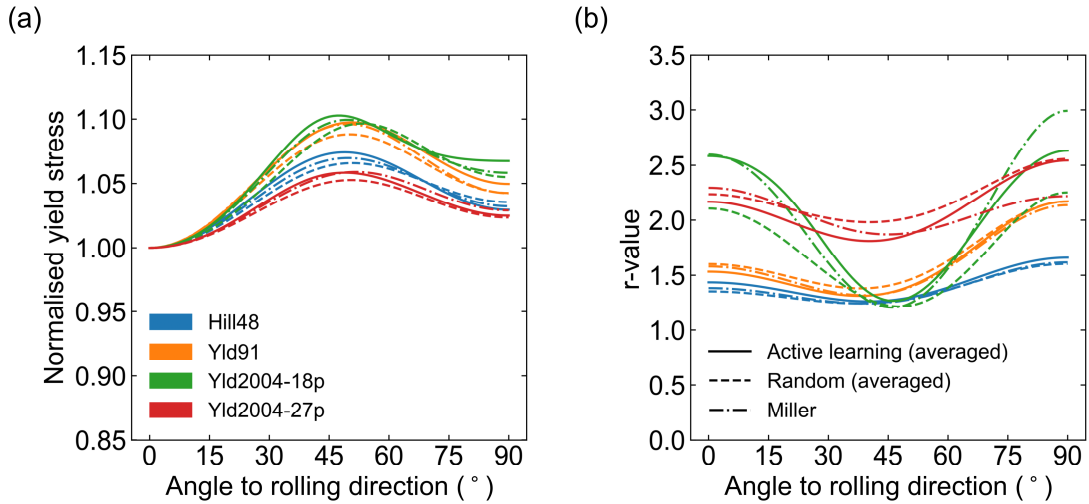

Fig. 13: (a) Normalised yield stresses and (b) the r-values with respect to RD for the Hill48, Yld91, Yld2004-18p and Yld2004-27p yield functions as identified by the active learning-based (solid line), random (dotted line) and Miller indices-based (dash-dotted line) sampling approaches. Yield surfaces identified by active learning-based and random sampling approaches were averaged by calculating the arithmetic mean of the five data sets.

3.5 In-plane anisotropy

To also evaluate the anisotropic yield surfaces with respect to the in-plane behaviour, 100 additional virtual experiments applying a plane stress state were carried out as stated in Section 2.2.4. The resulting yield points in Fig. 14 (a) were then utilised to identify the in-plane parameters of the Yld2004-18p and Yld2004-27p yield functions. As can been seen from Fig. 14 (b), both Yld2004-18p (in-plane) and Yld2004-27p (in-plane) match the yield points for $\sigma_{12} \approx 0$ with high accuracy. Also, the results of the uniaxial tensile tests at 0°, 15°, 30°, 45°, 60°, 75° and 90° with respect to RD in Fig. 14 (c) and (d), which were not part of the parameter identification, are captured accurately by both anisotropic yield functions.

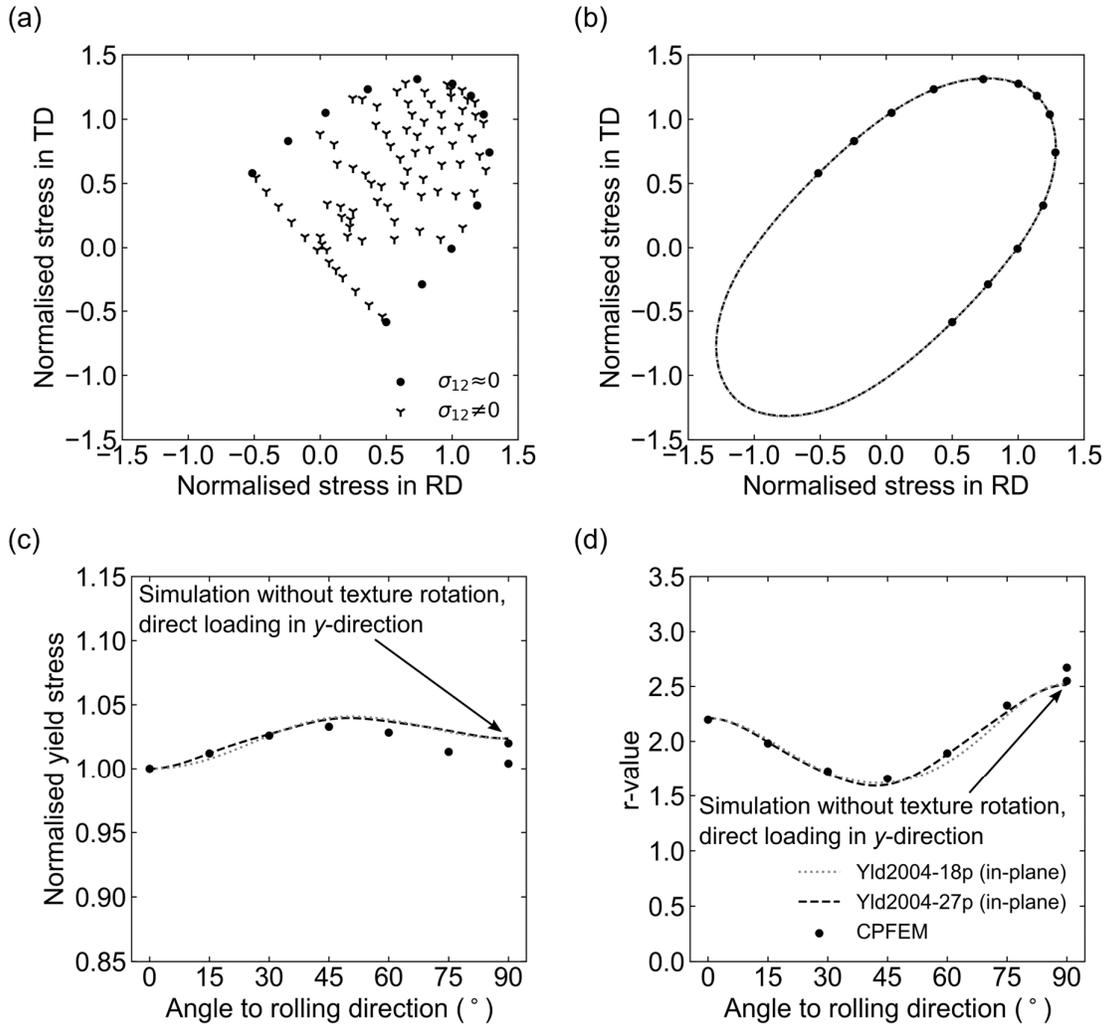

Fig. 14: Yld2004-18p and Yld2004-27p yield surfaces as identified by 100 virtual experiments applying a plane stress state: (a) 100 points on the yield surface as determined by the virtual experiments for the plane stress state, (b) normalised yield surface with respect to the RD-TD plane, (c) the normalised yield stresses as well as (d) the r-values with respect to RD.

The root-mean-square deviations (RMSD) of the Yld2004-18p (in-plane) and Yld2004-27p (in-plane) yield surfaces in Table 6 are relatively low as well, and the differences between the yield surfaces are rather negligible. As Yld2004-18p (in-plane) and Yld2004-27p (in-plane) represent the plastic anisotropy regarding the in-plane behaviour with nearly the same accuracy, only Yld2004-18p (in-plane) is illustrated in the following figures as a reference for the in-plane anisotropy.

Table 6: Root-mean-square deviations (RMSD) of the Yld2004-18p (in-plane) and Yld2004-27p (in-plane) yield surfaces with respect to the 100 virtual experiments applying a plane stress state.

|  | Yld2004-18p (in-plane) | Yld2004-27p (in-plane) |
| --- | --- | --- |
| RMSD | 0.0020 | 0.0017 |

Fig. 15 compares the anisotropic yield surfaces of Hill48-Active, Yld91-Active, Yld2004-18p-Active and Yld2004-27p-Active against the Yld2004-18p (in-plane) yield surface. As the yield surfaces of the three sampling methods in Fig. 12 and Fig. 13 are very similar, Fig. 15 only shows the results for the active learning-based sampling approach by way of example. Comparing the normalised yield surfaces in Fig. 15 (a) to (d), the best match for the Yld2004-18p (in-plane) yield surface is demonstrated by Yld2004-27p-Active. This good agreement between Yld2004-18p (in-plane) and Yld2004-27p-Active is also visible in the results of the normalised yield stresses and r-values in Fig. 15 (e) and (f).

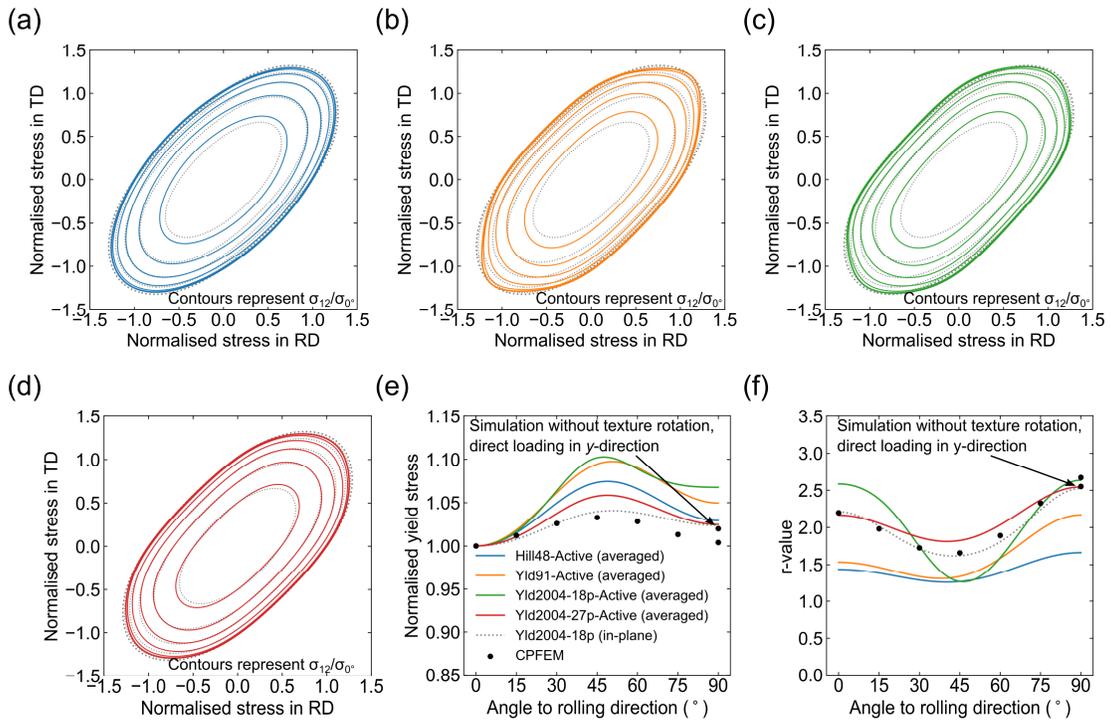

Fig. 15: Normalised yield stresses with respect to the RD-TD plane for the (a) Hill48-Active, (b) Yld91-Active, (c) Yld2004-18p-Active and (d) Yld2004-27p-Active yield surfaces as well as (e) the normalised yield stresses and (f) the r-values with respect to RD compared to the reference solution. The reference solution is given by Yld2004-18p (in-plane), whose parameters were determined by using virtual experiments applying a plane stress state. Yield surfaces identified by the active learning-based sampling approach were averaged by calculating the arithmetic mean of the five data sets. Normalised shear contours shown in increments of 0.1 from 0.0 to 0.5.

To further analyse the Hill48, Yld91, Yld2004-18p and Yld2004-27p yield surfaces with respect to the in-plane behaviour, the RMSD regarding the 100 virtual experiments applying a plane stress state was calculated for all three sampling methods. As can be seen from Table 7, the lowest RMSD is given by the Yld2004-27p yield surfaces regardless of the sampling method. The results of the RMSDs for the Hill48, Yld91 and Yld2004-18p yield surfaces are notably higher and of similar magnitude.

Table 7: RMSDs of the Hill48, Yld91, Yld2004-18p and Yld2004-27p yield surfaces with respect to the 100 virtual experiments for the plane stress state. RMSDs for the active learning-based and the random sampling approaches were averaged by calculating the arithmetic mean of the five data sets.

|  | Hill48 | Yld91 | Yld2004-18p | Yld2004-27p |
| --- | --- | --- | --- | --- |
| Active learning | 0.0317 | 0.0325 | 0.0350 | 0.0143 |
| Miller | 0.0310 | 0.0318 | 0.0336 | 0.0125 |
| Random | 0.0301 | 0.0272 | 0.0342 | 0.0151 |

3.6 Application to cup drawing simulations

To analyse the effect of the three sampling methods and the corresponding difference in the anisotropic yield surfaces in respect of sheet metal forming simulations, all anisotropic yield surfaces were used to predict the earing profile of a cylindrical cup drawing process via finite element simulations. It should be noted that this process is plane stress dominated, and thus the result for the Yld2004-18p (in-plane) yield surface serves as a reference. An example of a deep drawn cup as well as the resulting earing profiles are shown in Fig. 16 and Fig. 17 respectively. As with the results previously shown, one can observe that all three sampling methods lead to comparable earing profiles. For example, the earing profiles obtained by Hill48-Active, Hill48-Miller and Hill48-Random do not show much variation. In contrast, differences between the anisotropic yield functions, e.g. Yld91-Active and Yld2004-27p-Active, are more pronounced. Furthermore, the results obtained with the active learning-based sampling approach – each line corresponds to one data set – show less scatter in the resulting earing profiles compared to the random sampling approach. In addition, the results for the Yld2004-27p yield functions exhibit the best agreement with respect to the Yld2004-18p (in-plane) yield surface. This is consistent with the analysis of the normalised yield surfaces in Fig.15.

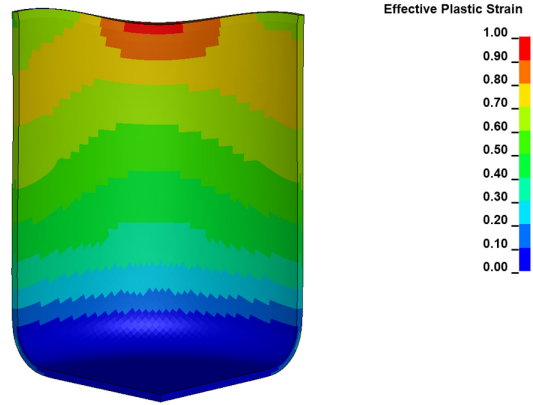

Fig. 16: Deep drawn cup using the Yld2004-27p-Active yield surface as identified by data set 3. Only a quarter of the blank was considered in the analysis due to the orthotropic material symmetry of the sheet.

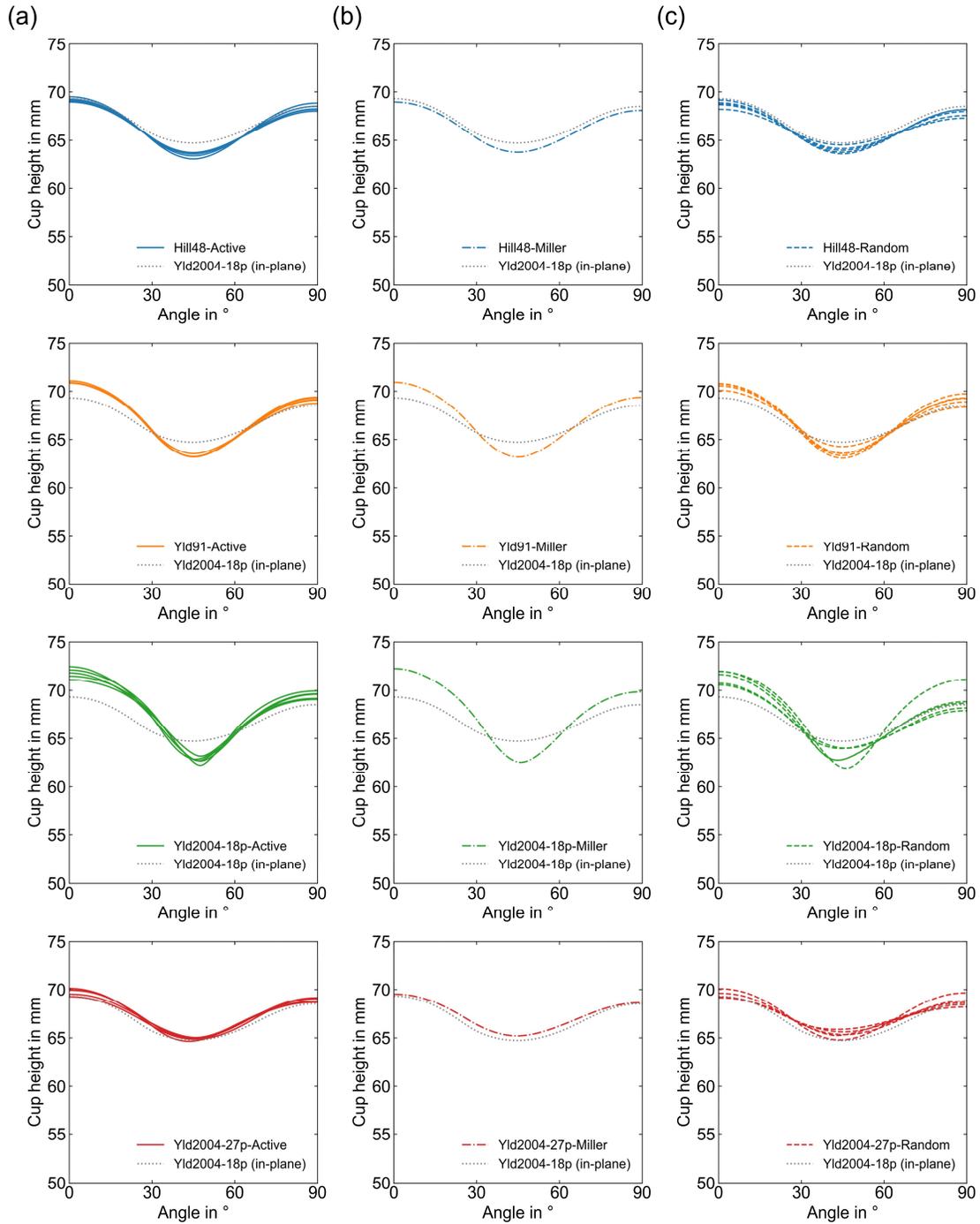

Fig. 17: Cup height as predicted by the Hill48, Yld91, Yld2004-18p and Yld2004-27p yield functions compared to the reference solution. Parameters were identified based on yield points generated by (a) the active learning-based sampling approach, (b) the random sampling approach and (c) the Miller indices-based sampling approach. The reference solution is given by Yld2004-18p (in-plane), whose parameters were determined by using virtual experiments applying a plane stress state.

## 4. Discussion

The discussion section is structured according to the main findings of this study. Firstly, the plastic anisotropy of DX56D deep drawing steel is discussed and the experimental results are compared with the literature. Secondly, the quality of the microstructure model used for representing DX56D deep drawing steel is analysed. Particular attention is given to the effect of the texture rotation on microstructure models with elongated grains. Thirdly, the results of the new active learning-based sampling approach for performing virtual experiments within the full stress state are assessed. This evaluation is based on the comparison with the results of the Miller indices-based and the random sampling approaches and focusses on the sampling efficiency as well as the reproducibility. Finally, the effect of different anisotropic yield functions for representing the full stress state is discussed.

### 4.1 Plastic anisotropy of DX56D deep drawing steel

Both the mechanical and crystallographic results in Section 3.1 demonstrate that the DX56D deep drawing steel under investigation has a strong anisotropic plastic material behaviour. The mechanical material properties generally correspond with the technical delivery conditions as given in DIN EN 10346 and are consistent with the mechanical and crystallographic results reported by Butz et al. (2019) for a different batch of DX56D. The results of the hydraulic bulge tests are in accordance with Sigvant et al. (2009). In addition, the results for the evaluation of the plastic anisotropy in Fig. 4 demonstrate that plastic anisotropy under uniaxial and biaxial loading conditions changes during deformation and only remains constant for strains above 10 to 15%. A similar behaviour for plastic anisotropy was observed by Volk et al. (2011) for a DX54D steel grade. According to Hill and Hutchinson (1992), this phenomenon whereby the plastic anisotropy, or rather the hardening behaviour, evolves under proportional monotonic loading conditions, is defined as differential work hardening. As differential work hardening is generally known in interstitial free (IF) and low-carbon steel grades, see Kuwabara et al. (1998) and Eyckens et al. (2015) for example, it would appear to be the most likely explanation for the differences observed with respect to the plastic anisotropy or hardening behaviour. In addition, physical mechanisms for differential work hardening include texture development and strain heterogeneity on the grain length scale, which is why crystal plasticity models are generally able to capture differential work hardening (Eyckens et al., 2015). The results in Fig. 9 (b) furnish evidence that the phenomenological crystal plasticity model used in this study can predict differential work hardening for commercial IF steels like DX56D with a reasonable degree of accuracy.

4.2 Quality of the microstructure model

The results in Section 3.2 demonstrate two important aspects regarding the quality of the microstructure model. First, due to the good match between experimentally and virtually determined yield stresses and r-values in Fig. 8, the microstructure model is well suited to represent the plastic anisotropy of DX56D deep drawing steel. Similar results have already been reported by Zhang et al. (2015), Zhang et al. (2016), Butz et al. (2019), Liu et al. (2020) and Engler and Aretz (2021), taking uniaxial tensile tests, plane strain tension tests and pure shear tests, as well as different aluminium alloys and steel grades into account. The results in Fig. 9 (a) demonstrate that virtual experiments can also predict the experimental biaxial flow stress of hydraulic bulge tests with a reasonable error of 3 to 4%. Second, the results in Fig. 8 illustrate that simulating uniaxial tensile tests in different directions with respect to RD on the basis of a texture rotation – as is often done in the literature, see Zhang et al. (2015) and Zhang et al. (2016) for instance – can cause a perceptible error for microstructures with elongated grains. This error was expected by the authors and is due to the fact that texture rotation excludes any effect of the grain morphology. Elongated grains with an aspect ratio of 1.6 were incorporated into the RVE for DX56D deep drawing steel. With the result for the normalised yield stress in TD in Fig. 8 (a), this led to an error of roughly 1.6% compared to the reference solution, with no rotation of the texture, and is still considered reasonable. As this error is expected to be governed by the size of the grain elongation, i.e. the error rises as the aspect ratio incorporated into the RVE increases, it should be taken into account for microstructures with higher grain elongation. Furthermore, all sampling methods were performed without rotating the texture, i.e. the prescribed deformation history was directly applied to the microstructure model. Therefore, the yield surfaces in Fig. 14 and Fig. 15 are in good agreement with the reference solution as obtained by a direct loading in the $y$-direction.

4.3 Active learning-based sampling of virtual experiments

The results presented in Sections 3.3, 3.4 and 3.6 demonstrate that the new active learning-based sampling approach is suitable for sampling virtual experiments within the full stress state in a data-efficient and reproduceable/reliable manner. Compared to the 402 virtual experiments performed for the Miller indices-based sampling approach, far fewer virtual experiments had to be performed with the new sampling approach to identify parameters for all anisotropic yield functions. For example, only 60 virtual experiments were necessary to identify the parameters of the Hill48 yield function. Nevertheless, differences in the Hill48-Active and Hill48-Miller yield surfaces in Fig. 12 and Fig. 13 are rather negligible and the two yield surfaces led to very similar earing profiles, as shown in Fig. 17. This advantage in the sampling efficiency also applied to the comparison with the random sampling approach considering the Hill48, Yld91, and Yld2004-18p yield functions. In contrast, differences in the sampling efficiency of the two sampling methods appear to be negligible for Yld2004-27p. This can be most likely explained by the high number of parameters of the Yld2004-27p yield function. Fig. 10 (a) to (c) illustrate that the active learning-based sampling approach improves the sampling of yield points when the number of virtual experiments performed is low. At the same time, advanced anisotropic yield functions like Yld2004-27p require a larger number of yield points for the parameter identification. It is assumed that the higher sampling efficiency of the active learning-based sampling approach diminishes with increasing number of yield points, so that almost no differences were observed for the Yld2004-27p function. However, it must be noted that the results for the active learning learning-based sampling approach in Fig. 10 show a smaller standard error than the random sampling approach. The implications of a lower standard error are also visible in the results of the cylindrical cup drawing simulation in Fig. 17. Compared to the earing profiles obtained with the anisotropic yield surfaces form the random sampling approach, those for the active learning-based sampling approach deviate less for the Hill48, Yld91, Yld2004-18p and Yld2004-27p yield functions. This is beneficial for the reproducibility and is particularly important with respect to a potential industrial application of this method. In conclusion, the active learning-based sampling approach is a data-efficient and reliable sampling method for the full stress state. Both findings concerning the efficiency as well as the reliability are consistent with the results of Wessel et al. (2021) when a related active learning approach was applied to the plane stress state. In addition, the findings are also consistent with Qu et al. (2023). Here, the concept of active learning was successfully applied to improve the training of data-driven constitutive models for granular materials. In the future, the efficiency of the active learning-based sampling approach can most likely be further improved. For instance, at present the parameter identification of the active learning-based sampling approach relies only on points on the initial yield surface. By taking the corresponding plastic strain ratios of the yield points into account, the number of virtual experiments can most likely be further reduced.

4.4 Anisotropic yield functions for the full stress state

The comparison between the four anisotropic yield surfaces Hill48, Yld91, Yld2004-18p and Yld2004-27p in Section 3.4 as well as the analysis of the in-plane anisotropy in Section 3.5 reveal two further important aspects with respect to virtual experiments and anisotropic yield functions for the full stress state: Firstly, as shown in Fig. 12 and Fig. 13, various sampling approaches with different sampling efficiencies can lead to very similar anisotropic yield surfaces – assuming enough yield points were sampled. Secondly, identifying parameters of anisotropic yield functions based on yield points sampled within the full stress state can lead to a degraded representation of the in-plane anisotropy. For example, both Yld2004-18p (in-plane) and Yld2004-27p (in-plane) in Fig. 14 are flexible enough to represent the plastic anisotropy of DX56D with respect to the in-plane behaviour. However, when taking the full stress state into account, only Yld2004-27p was able to represent the plastic anisotropy with respect to the in-plane behaviour with reasonable accuracy, as shown in Fig. 15. For Yld2004-18p, taking the out-of-plane anisotropy into consideration led to a degraded representation of the in-plane anisotropy, which is further verified by the results of the RMSD in Table 6. The negative implications of this degraded representation of the in-plane anisotropy for sheet metal forming simulations are further visualised by the results of the earing profiles in Figure 17. Whereas all Yld2004-18p yield surfaces exhibited a noticeable deviation from the reference solution Yld2004-18p (in-plane), the results of the Yld2004-27p yield surfaces demonstrated good agreement. This degraded representation of the in-plane anisotropy for the Yld2004-18p yield surfaces is caused by the mathematical formulation of this yield function. Yld2004-18p is based on two linear transformations of the deviatoric stress tensor. For this reason, material parameters associated with the in-plane and out-of-plane anisotropy are not independent, but rather coupled when considering the full stress state. As a consequence, the use of yield points sampled within the full stress state can result in a degraded representation of the in-plane anisotropy as was demonstrated by the results of Yld2004-18p. To improve the representation of the in-plane anisotropy when considering the full stress state, a sequential parameter identification might be sensible. For example, in-plane parameters are first identified using the results of virtual experiments applying a plane stress state. Second, out-of-plane parameters are calibrated based on yield points sampled within the full stress state. The same dependency between in-plane and out-of-plane material parameters also applies to the Yld2004-27p yield function, whose mathematical formulation is based on three linear transformations of the deviatoric stress tensor. However, because it has more material parameters, Yld2004-27p was flexible enough to accurately describe the plastic anisotropy of DX56D deep drawing steel with respect to both in-plane and out-of-plane behaviour.

## 5. Conclusions

This study introduces a new machine learning-based sampling approach based on active learning to perform virtual experiments with respect to the full stress state and analyses the effect of different sampling methods on the identification of parameters of anisotropy yield functions. To this end, virtual experiments were carried out for a DX56D deep drawing steel using the new active learning-based sampling approach and two state-of-the-art sampling methods taken from the literature by way of comparison. The benchmarking revealed that the new active learning-based sampling approach is beneficial for sampling points within the full stress state. Compared with the two state-of-the art sampling approaches taken from the literature, this new sampling method has a higher sampling efficiency. It also guarantees better reproducibility than the random sampling approach. Furthermore, the analysis of the Hill48, Yld91, Yld2004-18p and Yld2004-27p yield functions revealed that although these three sampling methods differ regarding their sampling efficiency, the resulting yield surfaces are comparable when enough yield points are sampled. It was also found that identifying parameters of anisotropic yield functions based on virtual experiments sampled within the full stress state can lead to a degraded representation of the in-plane anisotropy. This degradation was observed for the Yld2004-18p yield function and can generally occur for anisotropic yield functions whose in-plane and out-of-plane parameters are coupled. As a result, the representation of in-plane anisotropy must be carefully reviewed when the full stress state is considered. For DX56D deep drawing steel, Yld2004-27p was flexible enough to simultaneously describe the plastic anisotropy with respect to both the in-plane and out-of-plane behaviour with high accuracy.

**CRediT authorship contribution statement**

**Alexander Wessel:** Conceptualisation, Funding acquisition, Investigation, Methodology, Software, Writing – original draft, Visualisation. **Lukas Morand:** Methodology, Software, Writing – original draft, Visualisation. **Alexander Butz:** Funding acquisition, Writing – review & editing, Supervision. **Dirk Helm:** Funding acquisition, Writing – review & editing, Supervision. **Wolfram Volk:** Writing – review & editing, Supervision.

**Declaration of competing interest**

The authors declare that they have no known competing financial interests or personal relationships that could have appeared to influence the work reported in this paper.

**Acknowledgement**

The authors gratefully acknowledge funding from the Federal Ministry for Economic Affairs and Climate Action via the German Federation of Industrial Research Associations – AiF (Arbeitsgemeinschaft industrieller Forschungsvereinigungen e.V.) within the scope of the programme for Industrial Collective Research (Industrielle Gemeinschaftsforschung, IGF), grant numbers 19707 N and 21466 N, and the German Research Foundation (Deutsche Forschungsgemeinschaft, DFG), project number 415804944. The authors would like to thank Dr. Christoph Schmied and Dr. Tobias Erhart from DYNAmore GmbH for implementing the anisotropic yield function Yld2004-27p into LS-DYNA. A. Wessel thanks Valerie Scholes for proofreading of this article.

**Research data**

Research data to this article can be found online at http://dx.doi.org/10.24406/fordatis/225.
41